\documentclass[]{aa}  

\usepackage{graphicx}
\usepackage{xcolor}
\usepackage{txfonts}
\usepackage{natbib}
\usepackage{hyperref}
\hypersetup{
    colorlinks=true,
    citecolor=blue,
    linkcolor=blue,
    urlcolor=blue,
}

\makeatletter
\renewcommand*\aa@pageof{, page \thepage{} of \pageref*{LastPage}}
\makeatother

\newcommand{\edit}[1]{#1}
\newcommand{\remove}[1]{}

\begin{document}

\title{Atmospheric heating and magnetism driven by\\ $^{22}$Ne distillation in isolated white dwarfs}

\author{A.~F.~Lanza
\inst{1}
\and
N.~Z.\,Rui\inst{2}
\and
J.~Farihi\inst{3}
\and
J.~D.~Landstreet\inst{4,5}
\and
S.~Bagnulo\inst{4}
}

\institute{INAF-Osservatorio Astrofisico di Catania, Via S.~Sofia, 78 - I-95123 Catania, Italy\\
\email{antonino.lanza@inaf.it}
\and
TAPIR, California Institute of Technology, Pasadena, CA 91125, USA
\and
Department of Physics and Astronomy, University College London, London WC1E 6BT
\and
Armagh Observatory \& Planetarium, College Hill, Armagh BT61 9DG, UK
\and
University of Western Ontario, London, Ontario N6A 3K7, Canada
}

\titlerunning{Atmospheric heating in magnetic white dwarfs}
 
\abstract{The origin of atmospheric heating in the cool, magnetic white dwarf GD\,356 remains unsolved nearly 40 years after its discovery. This once idiosyncratic star with $T_{\rm eff}\approx7500$\,K, yet Balmer lines in Zeeman-split emission is now part of a growing class of white dwarfs exhibiting similar features, and which are tightly clustered in the HR diagram suggesting an intrinsic power source. This paper proposes that convective motions associated with an internal dynamo can power electric currents along magnetic field lines that heat the atmosphere via Ohmic dissipation. Such currents would require a dynamo \edit{driven by core $^{22}$Ne distillation}, and would further corroborate magnetic field generation in white dwarfs by this process. The model predicts that the heating will be highest near the magnetic poles, and virtually absent toward the equator, in agreement with observations. This picture is also consistent with the absence of X-ray or extreme ultraviolet emission, because the resistivity would decrease by several orders of magnitude at the typical coronal temperatures. \edit{The proposed model suggests that i) DAHe stars are mergers with enhanced $^{22}$Ne that enables distillation and may result in significant cooling delays; and ii) any mergers that distill neon will generate magnetism and chromospheres.  The predicted chromospheric emission is consistent with the two known massive DQe white dwarfs.
}}


\maketitle

\nolinenumbers 
\section{Introduction}

Magnetism in white dwarfs has been known for over half a century \citep{Kemp1970,Angel1971}, yet its origins remain somewhat elusive, despite recent progress. Broadly speaking, such magnetic fields are either remnants from dynamos during prior evolutionary stages (i.e.\ fossil fields) or sustained by some kind of currently active dynamo. While precise field formation mechanisms are difficult or impossible to infer on case-by-case bases, they are likely encoded within features of the magnetic white dwarf population as a whole.

The observed magnetic white dwarf population probably requires at least two distinct formation channels \citep{Bagnulo2022}. While magnetism in white dwarfs with $M\gtrsim1.1M_\odot$ almost certainly originates from dynamos that operate during their progenitor \edit{merger} \citep{Garcia-Berro2012,Bagnulo2022,Kiletal23} and can manifest within a few 100\,Myr, surface magnetic fields in $M\lesssim0.75M_\odot$ white dwarfs clearly arise later, generally after 2\,Gyr \citep{Bagnulo2021,Bagnulo2022}. Hypotheses regarding field formation in canonical mass white dwarfs must therefore account for the time delay for the onset of detectable fields. For example, fossil field hypotheses have suggested this delay could be the Ohmic diffusion time required for a buried core field to erupt at the white dwarf surface \citep{BlatmanGinzburg23}.

A promising alternative mechanism invokes a crystallization-driven dynamo to generate the required fields \citep{Isern2017}. In this scenario, a phase transition that occurs in carbon--oxygen core white dwarfs causes a composition-driven dynamo at the crystallization boundary. Notably, because the onset of crystallization requires sufficient white dwarf cooling, it may account for the observed time delay \citep{Bagnulo2021,Bagnulo2022,Amoretal23, Haretal23}. However, there are theoretical objections based on a possible insufficiency of kinetic energy for the dynamo \citep{Fuentesetal23,MontgomeryDunlap23}.

Remarkably, a subclass of magnetic white dwarfs (of spectral type DAHe\footnote{A degenerate star spectrum (D), with Balmer lines strongest (A), Zeeman-split (H), and in emission (e).}) exhibit Balmer emission lines in their spectra. This requires an atmospheric temperature inversion, that is, a chromosphere, that in turn requires some (as of yet unknown) heating mechanism.

The prototype is GD\,356, which was the sole example for over three decades \edit{\citep{Greenstein1985}}, and has been thoroughly investigated for signatures of accretion or a corona, and substellar companions down to the deuterium-burning limit \citep{Ferrario1997,Weisskopf2007,Wickramasinghe2010}. The lack of conventional, external actors led to a hypothesis where the atmospheric heating is caused by Ohmic dissipation in a current loop set up by an orbiting and conducting planet \citep[the unipolar inductor model;][]{Li1998}. In this model, similar to star--planet interactions for solar-type stars, the planet is magnetically connected to the white dwarf at a point (footprint) that is offset from the magnetic pole \citep[e.g.][]{Zarka2007,Lanza13,Kavanagh2023}. As the planet orbits, the heated footprint should vary its position on the stellar surface, as observed for the Jupiter--Io system \citep{Piddington1968,Goldreich1969}.

However, a growing body of observations has cast substantial doubt on the unipolar inductor model. GD\,356 has been studied in detail using time-series spectroscopy and spectropolarimetry over several rotation periods, as well as via multi-band, variable light curves that span several thousands of its 1.9\,h spin cycles, and including some theoretical considerations \citep{Walters2021}. First, all data are consistent with a single periodic modulation via stellar rotation, now including several years of {TESS} observations (N.~Walters 2024, private communication). Second, unlike solar-type stars that generate their own current sheets via winds, there is no source of magnetospheric ions for isolated white dwarfs. Third, the extreme centrifugal forces induced by rapid stellar rotation almost certainly inhibit current carriers in a unipolar inductor model. All evidence suggests the chromospheres are an intrinsic stage of evolution for (some or all) magnetic white dwarfs \citep{Walters2021}. 

In the past year, the understanding of the DAHe white dwarf population has been enhanced by roughly two dozen new discoveries \citep[e.g.][]{Reding2023,Manser2023}.  For example, these objects collectively have fields $B\gtrsim5$\,MG, \edit{somewhat higher} masses $M\simeq0.8M_\odot$, and \edit{relatively fast} spin periods between minutes and hours \citep{Gaensicke2020,Walters2021}. Most remarkably, all known DAHe stars cluster tightly around effective temperatures $T_{\rm eff}\simeq7500$\,K or, equivalently, a narrow range of cooling ages \citep{Gaensicke2020,Walters2021}. 

\begin{figure}
\includegraphics[width=\linewidth]{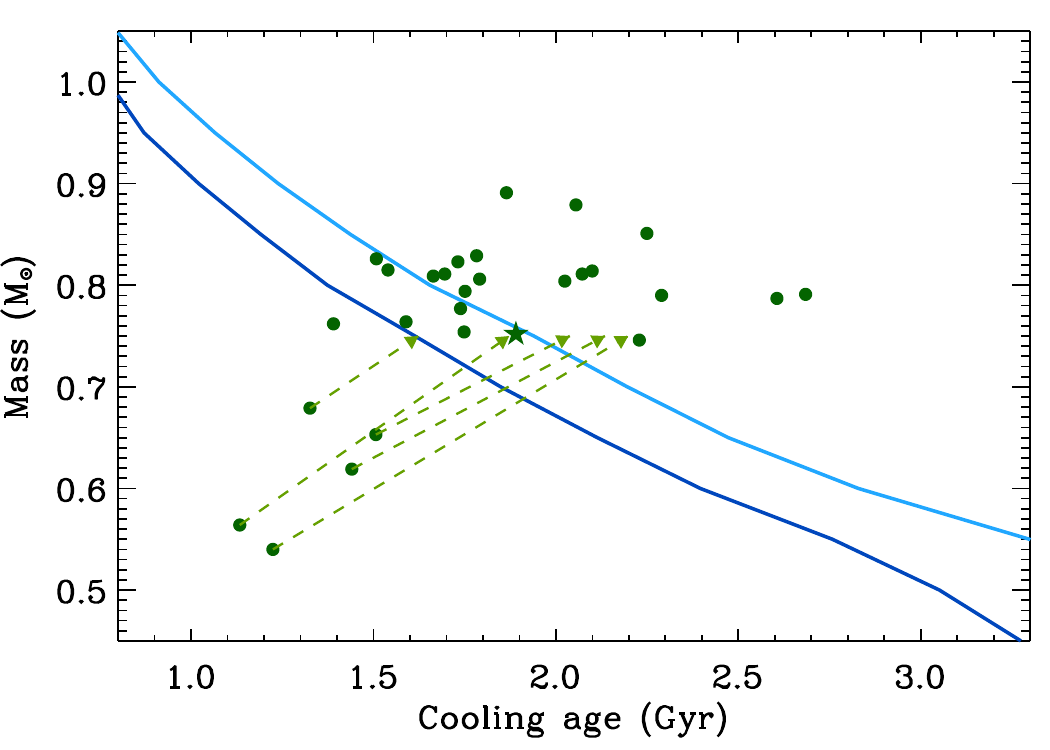}
\caption{Mass versus cooling age for currently known DAHe white dwarfs, plotted as dark green circles (with a star marking GD\,356).  The dark and light blue lines trace the first non-zero core crystallization fraction in hydrogen-rich white dwarfs for thin and thick envelopes, respectively \citep{Bedard2020}.  The masses and effective temperatures for all stars were taken from the {\em Gaia} EDR3 white dwarf catalog \citep{Gentile2021}.  While these adopted parameters are thus consistently derived, {\em Gaia} photometry alone does not provide a robust determination of effective temperature, where intrinsic photometric variability of DAHe stars, or missing model opacities may contribute to underestimated masses and cooling ages \citep[e.g.][]{Reding2020,Obrienetal24}.  For the lowest mass points in the diagram, the light green dashed lines trace their shift in cooling age, for the case where their masses are near the minimum value for the main cluster of stars.  Thus, it is plausible that all DAHe white dwarfs appear after the onset of crystallization.
\label{dahe_hrd}}
\end{figure}

All published DAHe white dwarfs are plotted as a function of estimated mass and cooling age in Fig.~\ref{dahe_hrd}. The main cluster of points is centered near 0.8\,M$_\odot$ and with a cooling age beyond the \edit{carbon--oxygen} crystallization front. While there are objects plotted at lower masses and hence younger ages for a given $T_{\rm eff}$, it is important to note that temperatures may have variable accuracy when derived by fitting photometric data for magnetic white dwarfs, owing to unaccounted for effects on continuum opacity \citep[see discussion in][]{Ferrario2015}. The problem may be exacerbated for DAHe stars that are photometrically variable at the few percent level, with a strong wavelength dependence \citep{Farihi2023}. Furthermore, it is well known that {\em Gaia} photometry has limited color information from only two non-overlapping filter bandpasses, and for cooler white dwarfs it appears that masses are often under-predicted by model fits to this photometry \citep{Obrienetal24}. Uncertainties in temperature propagate directly into errors in mass using {\em Gaia} parallaxes, and in turn cooling age. In any case, the clustering in cooling age further suggests that some kind of intrinsic heating mechanism exists.

There has been limited but important prior theory on intrinsic heating mechanisms for white dwarf atmospheres. It has been proposed that acoustic and magnetohydrodynamic waves can propagate into white dwarf atmospheres. However, damping processes are strong and possibly overwhelming, implying that any detectable oscillations will be no more than a few percent in luminosity and last no longer than a few seconds \citep{Musielak1987,Musielak1989,Musielak2005}.
The strong magnetic fields characteristic of the DAHe class will hinder convective motions and thus alter energy transport and structure of these stars, implying that conventional white dwarf atmospheres are unlikely to result in the observed heating. Therefore, a different mechanism of chromospheric heating is needed to account for the DAHe phenomenon. 

\begin{figure*}
\includegraphics[width=\linewidth]{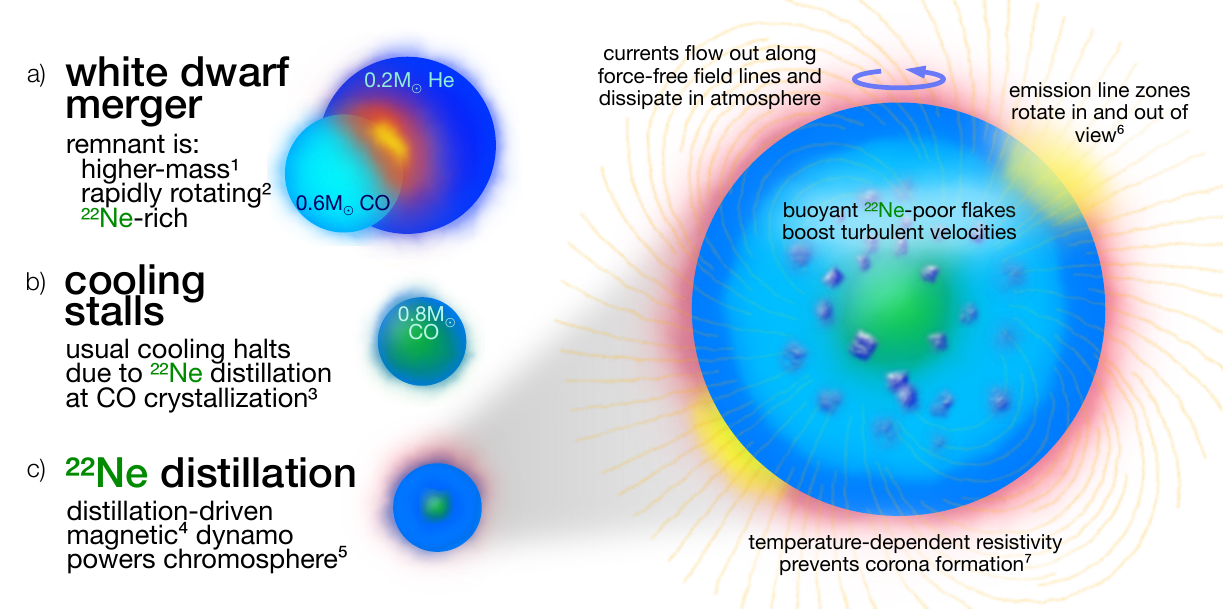}
\caption{\edit{Illustrative summary of the model advanced in this work.  (\textit{a})  DAHe white dwarf created by a stellar merger, such as between a carbon--oxygen and helium-core white dwarf \citep[as shown by, e.g.][]{clayton2007very,menon2013reproducing} or subgiant. (\textit{b})  Resulting $^{22}$Ne-rich remnant cools until the onset of carbon--oxygen crystallization and $^{22}$Ne distillation, which halt the cooling near $T_{\mathrm{eff}}\approx7500\,\mathrm{K}$. (\textit{c})  Distillation also provides sufficient energy to sustain a strong magnetic dynamo that powers the characteristic chromospheric emission through Ohmic heating.   Superscript numbers indicate observed properties of DAHe stars: $^{1}$typical masses $M\simeq0.8M_\odot$, $^{2}$rapid rotation, $^{3}$similar $T_{\rm eff}\approx7500\,\mathrm{K}$ near that of carbon--oxygen crystallization, $^{4}$magnetic fields $B\gtrsim5$\,MG, $^{5}$Balmer lines in emission, $^{6}$rotational light curve modulation, and $^{7}$lack of coronal emission in the prototype.}
\label{cartoon}}
\end{figure*}

\edit{In this work, we propose that chromospheric emission in DAHe white dwarfs is powered by Ohmic heating associated with electric currents produced by a $^{22}$Ne distillation-driven dynamo during crystallization. We argue that these stars are likely merger remnants, consistent with their systematically high masses and rotation rates. The progenitor merger events can likely enhance the $^{22}$Ne abundances in DAHe stars to the level needed to sustain strong, long-lasting dynamo action \citep{clayton2007very,menon2013reproducing,Shenetal23}. This scenario is essentially a lower-mass analogue of the $^{22}$Ne-distillation mechanism proposed to stall the cooling of ultramassive Q-branch white dwarfs \citep{Isernetal91,Segretain96,Blouinetal21,Bedardetal24,Salarisetal24}, and, if correct, establishes continuity between these two fascinating classes of objects.  Our model and its relationship to observations are summarized in Fig.~\ref{cartoon}.}

This work is organized as follows.  In Sect.~\ref{model}, we outline our model, where the generation of magnetic fields and electric currents by a hydromagnetic dynamo is presented in Sect.~\ref{dynamo_model}, while convective motions, that provide the basic ingredients for the dynamo action, are discussed in Sect.~\ref{wd_conv}.  The magnetic field diffusion from the dynamo shell to the surface is discussed in Sect.~\ref{field_diffusion}.  The dissipation of the electric currents in the atmosphere is treated in Sect.~\ref{field_dissipation}. We summarize our results, \edit{discuss immediate implications, and present considerations for future work} in Sect.~\ref{conclusions}.

\section{Model}
\label{model}

\subsection{Dynamo \edit{action} in white dwarf interiors}
\label{dynamo_model}

One of the proposed mechanisms for the origin of white dwarf magnetic fields is a hydromagnetic dynamo that takes its energy from the convective motions produced by core crystallization. Such a dynamo is similar to that operating in the outer fluid core of the Earth, which surrounds its inner solid core, because in white dwarfs the inner crystallized core is surrounded by a convective fluid shell \citep[][]{Isern2017}. The fast rotation observed in DAHe white dwarfs should impart a remarkable helicity to the convective flows in the outer shell because of the strong Coriolis force. Therefore, such a white dwarf dynamo is likely to be of the so-called $\alpha^{2}$ type, similar to those in the core of the Earth or in Jupiter, leading to a stationary magnetic field \citep{Moffatt1978}. 

The mean-field induction equation giving the magnetic field generated by an $\alpha^{2}$ dynamo is
\begin{equation}
    \frac{\partial {\vec B}}{\partial t} = \nabla \times \left( \alpha {\vec B} - \eta_{\rm t} \nabla \times {\vec B}\right),
    \label{dynamo_eq}
\end{equation}
where ${\vec B}$ is the mean magnetic field, $t$ the time, $\alpha$ the parameter specifying the $\alpha$-effect in the dynamo, and $\eta_{\rm t}$ the turbulent magnetic diffusivity. The boundary conditions in the stellar dynamo are the continuity of the normal components of the magnetic field and of the current density across the boundaries of the dynamo shell, while in the case of the geodynamo, the field is assumed to be potential outside the shell because the Earth's mantle behaves as an electric insulator \citep[see, e.g.][]{GlatzmaierRoberts95}.

While $\alpha$ is a tensor, for simplicity we assume it to be symmetric and isotropic, thus reducing it to a scalar. Because $\alpha$ is related to the helicity of the turbulent flow, it changes sign from the northern to the southern hemisphere of the star, and is a function of the latitude because the Coriolis force depends on latitude.  The simplest expression one can adopt is $\alpha \propto \cos \theta$, where $\theta$ is the colatitude measured from the north pole. The turbulent diffusivity is also a tensor, but for the sake of simplicity, we assume it to be a scalar approximated as 
\begin{equation}
\eta_{\rm t} \sim \frac{1}{3} \, \ell \, \varv_{\rm t}, 
\label{etat}
\end{equation}
where $\ell$ is the characteristic length scale of the turbulent convective motions, while $\varv_{\rm t}$ is their mean velocity \citep[see, for example,][]{Charbonneau2020}. 

The simplest stationary mean magnetic field generated by the hydromagnetic dynamo described by Eq.~\eqref{dynamo_eq} has the expression
\begin{equation}
    {\vec B} =  \left( \frac{\eta_{\rm t}}{\alpha}\right)  \, \nabla \times {\vec B} \equiv \beta^{-1} \nabla \times {\vec B},
    \label{fff_1}
\end{equation}
which is the expression of a force-free field with the force-free parameter $\beta = \alpha/\eta_{\rm t}$. \citet{Chandrasekhar1981} showed that a non-vanishing magnetic field cannot be force-free everywhere, and thus the field must be stressed somewhere, with pressure gradients or the local gravity balancing the Lorentz force. We assume that the field is stressed at the base of the dynamo layer because the density of the fluid is larger there, and the convective velocity should not depend strongly on depth (cf.\ Sect.~\ref{wd_conv}). This implies that the main field amplification process and the associated magnetic stresses are concentrated close to the base of the layer. This assumption should also be valid in the case of a non-convective dynamo, for example a Tayler--Spruit dynamo \citep{Spruit02}, which has recently been suggested to work during crystallization \citep{MontgomeryDunlap23}.

Our dynamo magnetic field is not necessarily axisymmetric.  Mean-field $\alpha^{2}$ dynamos based solely on the $\alpha$ effect admit non-axisymmetric dynamo solutions both in the linear and non-linear regimes, even if $\alpha$ is independent of the longitude \citep[e.g.][]{Ruediger80,Radleretal90,RuedigerElstner94,Moss05}. Specifically, a linear combination of modes with degree $l=1$ and azimuthal orders $m=-1,0,1$ yields a tilted dipole that can roughly describe the magnetic field of the Earth or Jupiter \citep[][Sect.~6]{Stevenson10}. We assume the same large-scale geometry can be produced in the fields of white dwarfs generated by our dynamo model, although we do not compute a full solution of Eq.~\eqref{dynamo_eq} as a linear combination of individual dynamo modes and instead limit ourselves to the simple illustrative expression given by Eq.~\eqref{fff_1}. We note that a tilted-dipole geometry is capable of accounting for the observed rotational modulation of the stellar flux \citep{Reding2020}, if the emission (or absorption) is concentrated around the poles of the field (see Sect.~\ref{field_dissipation}). 

In the domain where the field is force-free, the magnetic field vector $\vec B$ is parallel to the electric current density $\vec J = (\nabla \times {\vec B})/\mu$, where $\mu$ is the magnetic permeability of the vacuum. The parameter $\beta$ is constant along each magnetic field line as can be seen by taking the divergence of both sides of Eq.~\eqref{fff_1}, giving ${\vec B} \cdot \nabla \beta = 0$ because $\nabla \cdot {\vec B}=0$.  We note, however, that $\beta$ is distinct along different magnetic field lines because $\alpha$ and $\eta_{\rm t}$ are functions of the colatitude and the radial distance from the centre of the star. 

An approximate expression for the order of magnitude of $\beta$ can be obtained by noticing that the order of magnitude of $\alpha \sim \ell \, \Omega$, where $\Omega$ is the stellar spin angular velocity. Taking into account Eq.~\eqref{etat}, we find
\begin{equation}
    \beta \sim \frac{\alpha}{\eta_{\rm t}} \sim \frac{3\Omega \cos \theta}{\varv_{\rm t}},
    \label{beta_exp}
\end{equation}
which is independent of the (poorly known) mixing length $\ell$. When $\varv_{\rm t}$ goes to zero, the value of $\beta$ diverges, but the electric current density, that is proportional to $\beta {\vec B}$, remains finite provided that the magnetic field strength is proportional to $\varv_{\rm t}$. This is indeed the case when an $\alpha^{2}$-dynamo reaches equipartition between the kinetic energy density of the turbulent motions and the magnetic energy density, viz.\ $B^{2}/\mu \sim \rho \varv_{\rm t}^{2}$, where $\rho$ is the density of the plasma.

\subsection{\edit{Compositionally driven convection and the role of $^{22}$Ne distillation}}
\label{wd_conv}

The convective velocity in the dynamo fluid shell outside the crystallized core is still a subject of debate. The interior convective shell is stable to ordinary thermal convection, so convective motions are accelerated by the gradient of the molecular weight produced by oxygen crystallization. \edit{In the case of a pure carbon--oxygen mixture, solid oxygen is denser than the fluid from which it separates}, and thus it sinks and forms a solid core at the centre of the star, above which the carbon--oxygen mixture remains in the liquid phase. 

\edit{In recent years, the convective velocities $\varv_{\mathrm{t}}$ associated with canonical carbon--oxygen crystallization have been steadily revised downward. \citet{Isern2017} first derived an upper limit $\approx35$\,km\,s$^{-1}$, but this was later refined downward to $\simeq1$\,m\,s$^{-1}$ by \citet{Ginzburgetal22_WD_dynamo} when considering the mixing of buoyantly rising material, and even further by \citet{Fuentesetal23} when accounting for the stabilizing effect of the thermal stratification.} In white dwarfs with a rotation period of the order of one hour, the Coriolis force increases the velocity of the compositional convection in the regime considered by \citet{Fuentesetal23}, but even in that case the velocity does not exceed $10^{-4}$\,m\,s$^{-1}$ \citep[see][for a similar conclusion]{MontgomeryDunlap23}. Owing to the large dimensions of the convective eddies (comparable with the pressure scale height $\sim 10^{6}$\,m) and the small magnetic diffusivity in a degenerate plasma, the dynamo number is still supercritical and dynamo action can occur. However, the generated magnetic field strengths are at most of the order of several hundred of gauss, orders of magnitude smaller than required to account for the observations. This happens even when we assume that the Lorentz force reaches equipartition with the Coriolis force, as expected in fast-rotating, highly supercritical $\alpha^{2}$ dynamos, because the convective velocity is extremely small \citep[cf.\ sect.~3.2 of][]{Ginzburgetal22_WD_dynamo}.

\edit{Furthermore, while a dynamo generated by the onset of crystallization may be able to produce the $1-100$\,MG fields characteristic of magnetic white dwarfs \citep{Castro-Tapiaetal24,Fuentesetal24}, the associated convection is only sufficient for a small fraction of a white dwarf lifetime ($\sim10$\,Myr for $M=0.9M_{\odot}$). It is therefore unlikely to power the chromospheric emission in DAHe stars, which may represent $\approx14$\% of magnetic white dwarfs at the relevant $T_{\mathrm{eff}}$ \citep{Manser2023}.  If DAHe stars are merger remnants, a merger-driven dynamo may also produce a strong field \citep{schneider2019stellar}, but would not result in a lasting dynamo which could power the chromosphere.}

To overcome the low velocities predicted by the model of \citet{Fuentesetal23} for the case of compositional convection, \edit{we conjecture that motions associated with the distillation of $^{22}$Ne (originally proposed by \citealt{Isernetal91}) can produce sufficiently fast flows.   To be more specific, we consider the case of a white dwarf with central composition mass-fractions $X(^{16}\mathrm{O})=0.6$, $X(^{12}\mathrm{C})=0.365$, and $X(^{22}\mathrm{Ne})=0.035$ \citep[case (a) discussed by][]{Blouinetal21}. Such a $^{22}$Ne-rich composition is possible when the progenitor of a carbon--oxygen white dwarf formed in a particularly $\alpha$-rich environment or when it resulted from a \edit{merger} \citep[see][]{clayton2007very,menon2013reproducing,Shenetal23, Salarisetal24}, where corresponding evolutionary models have largely reproduced the overdensity of stars in the Q branch of the white dwarf HR diagram \citep[cf.][]{Tremblayetal19,Chengetal2019,Bedardetal24}. Indeed, the relatively high mass and fast rotation that characterize DAHe stars can be easily accounted for if they are merger remnants. In our model, it is assumed the conditions for $^{22}$Ne distillation are met, and leave detailed studies of specific merger scenarios to future work.}

\edit{We estimate the size and the rising velocity of the solid flakes formed by the $^{22}$Ne distillation in Appendix~\ref{Appendix_A}, finding these to be of the order of $10^{-4}$\,m and $(5-10)$\,m\,s$^{-1}$, respectively.} During its rise, a solid flake produces a turbulent wake that extends across the full vertical depth it spans. The slow diffusion of these turbulent wakes under the action of the small kinematic viscosity $\nu$ in a white dwarf interior \citep[$\nu \sim 3\times 10^{-6}$ m$^{2}$\,s$^{-1}$;][]{Isern2017} produces a turbulent velocity field remarkably faster than that due to the compositional convection alone.

The mean-field dynamo equations \citep{Moffatt1978,Charbonneau2020} can be non-dimensionalized with different choices for the unit of length. Here, we propose that the appropriate length scale is $\zeta l_{\rm w}$ instead of the star radius, where $\zeta$ is the volume filling factor of the flake turbulent wakes and $l_{\rm w}$ is their average vertical extension. \edit{The value of $l_{\rm w} \sim (|\Delta \rho|/\rho) H_{\rho}$, where $H_{\rho} \sim 3 \times 10^{6}$\,m is the density scale height at the surface of the core of a 0.9\,M$_{\odot}$ white dwarf \citep[cf.][table~1]{Fuentesetal24}.} Therefore, we conjecture that the dynamo number can be defined as $D = \zeta \alpha l_{\rm w}/\eta_{\rm t}$. Considering a white dwarf with $\Omega = 10^{-3} $\,s$^{-1}$, \edit{a conservative turbulent velocity $\varv_{\rm t} \sim 1$\,m\,s$^{-1}$}, and a length of the wakes of $l_{\rm w} \sim 3 \times 10^{5}$\,m, \edit{we find a supercritical dynamo number $D \sim 10$ for a volume filling factor $\zeta \sim 0.03$. The geometry of the most easily excited magnetic field mode is that of a dipole with a length scale comparable with the star radius \citep[cf.][Table~1]{Radleretal90}. } In such a dynamo, the magnetic field strength attained when the Lorentz force reaches equipartition with the Coriolis force is of the order of 10\,MG for a rotation period of 1~h and turbulent velocities of $\sim 1$\,m\,s$^{-1}$ \citep[cf.\ sect.~3.2 of][]{Ginzburgetal22_WD_dynamo}.

\subsection{Diffusion of the magnetic field to the stellar surface}
\label{field_diffusion}

In our simplified model, the mean field is force-free, that is, the Lorentz force vanishes. Therefore, the magnetic field does not affect the white dwarf equilibrium structure and stratification, which are instead set by a balance between pressure and gravity only. This is also true in the outer layers of the white dwarf, if the surface field is sufficiently strong (roughly MG) to halt near-surface convection. The vanishing Lorentz force also implies that magnetohydrodynamic forces such as magnetic pressure or magnetic buoyancy cannot bring magnetic fields to the surface (as in the Sun or sunlike stars). The strong field also quenches any plasma flow, thus preventing any transport by advection. Only magnetic diffusion can transport the field from the dynamo region to the stellar surface.

\citet{Ginzburgetal22_WD_dynamo} computed the magnetic diffusion timescale for two white dwarfs of masses 0.6 and 0.8\,M$_{\odot}$ finding that it becomes drastically reduced when the convection zone reaches the lower boundary of the helium envelope. At that point, the diffusion time from the dynamo shell to the atmosphere becomes of the order 0.1\,Gyr for 0.8\,M$_{\odot}$, and 1\,Gyr for 0.6\,M$_{\odot}$. A more detailed analysis by \citet{BlatmanGinzburg23} confirms such timescales, predicting a diffusion timescale of $\sim 0.6$\,Gyr shortly after the field breaks out at the surface of a 0.8\,M$_{\odot}$ white dwarf, which occurs when the star has a cooling age of $\sim 3$\,Gyr. 

An analysis of the diffusion is presented in Appendix~\ref{app_b_diff}, where we show that the field remains approximately force-free during the process. In such a regime, the field and the associated currents flowing along the field lines diffuse to the surface of the star reaching its photosphere. Those photospheric currents are responsible for the DAHe phenomenology because they produce a strong Ohmic heating in the atmosphere (cf. Sect.~\ref{field_dissipation}). 

Crucially, the magnetic field lines remain connected to the dynamo shell throughout the process of diffusing.  As a consequence, the value of the force-free parameter $\beta$ along each magnetic field line in the stellar atmosphere is set inside the dynamo region, because the force-free regime extends from the top of the dynamo shell to the atmosphere of the star.  The disconnection of the surface field from the dynamo region can occur only on the global diffusion timescale of the field itself, that is, on the order of 10\,Gyr, owing to its large spatial scale, $L$, that is comparable with the stellar radius.  In a white dwarf of 0.8\,M$_{\odot}$, the thickness of the layer separating the convection zone from the surface, $d_{\rm s} \sim 0.2\, R_{\rm s} < L $, and becomes shorter as the star cools \citep[cf.\ Fig.~1 of][]{BlatmanGinzburg23}.  

Our scenario aims to reproduce a steady atmospheric heating consistent with observations of the DAHe phenomenon. It therefore requires that the force-free fields generated by the dynamo are stable.  Unstable fields would be destroyed on the Alfv\'en \edit{timescale}. \citet{Molodensky74} found that force-free fields are stable to perturbations both in the photosphere \citep[due to the expected small-scale motions, e.g.][]{Tremblayetal15} as well as above the photosphere (such as those produced by hydromagnetic waves) as long as the typical length scale of such perturbations $l_{\rm pert}\ll\beta^{-1}$. Because $\beta^{-1}>d_{\rm s}\sim 0.2\, R_{\rm s}$ corresponds to the length scale of a global field in the present regime, this condition is certainly satisfied (see Sect.~\ref{field_dissipation} for a realistic estimate of $\beta$). It is interesting to note that small scale photospheric motions can lead to the formation of tangential discontinuities in the atmospheric magnetic fields that are sites of strong current concentration and heating by Ohmic dissipation \citep[see][for a recent review]{Low23}.

\subsection{\edit{Chromospheric heating power estimate}}
\label{field_dissipation}

In the absence of external energy sources, electric currents flowing parallel to the field in the atmosphere produce a steady dissipation of magnetic energy until the field reaches a potential configuration with $\beta=0$ after which there is no further dissipation (cf.\ Eq.~\ref{induction_eq}). However, \edit{even in the presence of strong atmospheric dissipation,} such a potential field state cannot be \edit{fully} reached while the hydromagnetic dynamo is in operation and the field lines remain connected with the dynamo shell.  In this regime, the dissipated electric energy is ultimately provided by the interior dynamo.

\edit{To calibrate our subsequent estimates, we supply some characteristic scales in this problem.}  For the DAHe prototype GD\,356, $R_{\rm s} = 0.01$\,R$_{\odot}$, $\Omega \sim 10^{-3}$\,s$^{-1}$, and $B \sim 10$\,MG. This radius is representative of carbon--oxygen white dwarfs in general, but the rotation rates of DAHe stars lie in the range $10^{-4}$--$10^{-2}$\,s$^{-1}$.  While the field strengths of DAHe stars are not tightly constrained (only GD\,356 has been monitored through its rotational cycle), they are estimated to be between 5 and 50\,MG \citep{Manser2023}, consistent with a disc-averaged mean field $\sim10$\,MG.  The convective velocity $\varv_{\rm t}$ is estimated to be $\varv_{\rm t} \sim 5-10$\,m\,s$^{-1}$ (see Sect.~\ref{wd_conv}). 

The dissipation per unit volume can be computed from the induction equation assuming that no motion occurs in the atmospheric plasma, so that 
\begin{equation}
\frac{\partial {\vec B}}{\partial t} = - \nabla \times \left( \eta \nabla \times {\vec B}\right),
\label{induction_eq}
\end{equation}
where $\eta$ is the microscopic  magnetic diffusivity in the atmosphere \citep{Spitzer62}, since turbulence and convection are strongly suppressed by the magnetic field \citep[e.g.][]{Saumon2022}. 

The magnetic diffusivity in the non-degenerate plasma of the stellar atmosphere is given by the usual Spitzer expression \citep[e.g.][]{Meyer1974}
\begin{equation}
    \eta = 3 \times 10^{8}  \, T^{-3/2} \mbox{ m$^{2}$\,s$^{-1}$},
    \label{eta_value}
\end{equation}
where $T$ is the plasma temperature in Kelvin. We discuss the suitability of such an expression in Appendix~\ref{magn_diff}. The greatest dissipation is reached in the atmosphere where $T$ is minimum. In the case of GD~356 and DAHe white dwarfs, $T\approx7500$\,K, \edit{so that} $\eta \approx 460$\,m$^{2}$\,s$^{-1}$. 

Taking the scalar product of Eq.~\eqref{induction_eq} by $\vec B$ and making use of the relationship for two generic vector fields $\vec u$ and $\vec w$, $\nabla \cdot ({\vec u} \times {\vec w}) = {\vec w} \cdot (\nabla \times {\vec u}) - {\vec u} \cdot (\nabla \times {\vec w})$, we obtain
\begin{equation}
    {\vec B} \cdot \frac{\partial {\vec B}}{\partial t} = \nabla \cdot \left[ \eta {\vec B} \times (\nabla \times {\vec B}) \right] - \eta (\nabla \times {\vec B})^{2},
    \label{energy_dot0}
\end{equation}
which can be reduced to
\begin{equation}
     \frac{\partial}{\partial t} \left( \frac{B^{2}}{2\mu}\right)=  - \frac{\eta}{\mu} (\nabla \times {\vec B})^{2},
     \label{energy_dot1}
\end{equation}
where we made use of the force-free configuration of the field ${\vec B}$ (cf. Eq.~\ref{fff_1}) to eliminate the first term on the right-hand side of Eq.~\eqref{energy_dot0}.

\edit{During the period of time when the magnetic field has not yet diffused to the surface, $\beta$ is given by Eq.~\eqref{beta_exp}. However, once the field has broken through, efficient Ohmic dissipation in the atmosphere globally affects the dynamo field as a boundary condition.  At this time (corresponding to the onset of the DAHe phenomenon), atmospheric diffusion modifies $\beta$ along an entire field line and sets the value of the current dissipation rate in the photosphere.}

\edit{As soon as the magnetic field reaches the surface, the values of $\beta$ characterizing the field are modified essentially instantaneously.
To see this, we recast Eq.~\eqref{energy_dot1} in the simpler form (making use of Eq.~\eqref{fff_1}):
\begin{equation}
    \frac{\partial}{\partial t} \left( \frac{B^{2}}{2\mu}\right)=  - 2 \eta \,\beta^{2} \, \left(\frac{B^{2}}{2\mu} \right).
    \label{diss_power}
\end{equation}}

\edit{For the initial values of $\beta$ given by Eq.~\eqref{beta_exp},}
\begin{equation}
    \frac{\partial}{\partial t} \left( \frac{B^{2}}{2\mu}\right) = - 2\eta \left( \frac{3 \Omega}{\varv_{\rm t}} \right)^2  \, \left(\frac{B^{2}}{2\mu} \right) \cos^{2} \theta \sim \frac{B^2/2\mu}{\tau_{\mathrm{diss}}},
\end{equation}
\edit{Ohmic dissipation therefore quickly relaxes the global field on a timescale}
\begin{equation}
    \tau_{\mathrm{diss}} \sim \frac{1}{18\eta}\left(\frac{\varv_{t}}{\Omega}\right)^2 \sim 1.5 \times 10^{4}\,\mathrm{s}
\end{equation}
\edit{when assuming a dynamo convective velocity $\varv_{\rm t} \sim 10$\,m\,s$^{-1}$, $\eta = 460$\,m$^{2}$\,s$^{-1}$, and $\Omega = 10^{-3}$\,s$^{-1}$.}

\edit{We argue that the resistive heating power experienced by the chromosphere is set by the rate at which the dynamo can amplify the magnetic field.  The associated timescale is comparable to the diffusion timescale}
\begin{equation}
    \tau_{\rm diff} \sim L^{2}_{\rm dyn}/\eta_{\rm t}
\end{equation}
across the dynamo region, with extent $L_{\rm dyn} \lesssim R_{\mathrm{s}}$ \citep{brandenburg2001inverse}.  In other words, $P_{\mathrm{diss}}$ is ultimately limited by the rate at which the dynamo can replenish the non-potential component of the field in the first place.

\edit{$P_{\mathrm{diss}}$ can be found by integrating $B^2/(2\mu\tau_{\mathrm{diss}})$ over the atmosphere.} The pressure scale height in the atmosphere of a white dwarf with effective temperature 7500\,K is $h\sim100$\,m, while the density near the photosphere (optical depth $\tau\sim 1$ for $\lambda\approx5000$~\AA) is approximately $10^{-2}$\,kg\,m$^{-3}$ \citep[e.g.][]{Saumon2022}. Roughly estimating that the H$\alpha$ line forms over an isothermal region of vertical extent $H\sim10h\sim1\,\mathrm{km}$ above the photosphere, the density at the top of the chromosphere is $\approx 4.5 \times 10^{-7}$\,kg\,m$^{-3}$, comparable with the density of the solar photosphere at its temperature minimum.

Assuming $L_{\rm dyn} \ga 3 \times 10^{6}$\,m, $l_{\rm w} \sim 3 \times 10^{5}$\,m, and $\varv_{\rm t} \sim 5-10$\,m\,s$^{-1}$ (to estimate $\eta_{\rm t}$ through Eq.~\ref{etat}), we find that $\tau_{\rm heat}\sim\tau_{\rm diff}\ga 10^{7}$\,s.  \edit{Finally, this implies a heating rate}
\begin{equation} \label{heatingrate}
    P_{\mathrm{diss}} \sim \frac{2\pi}{\mu}\frac{B^2R_{\rm s}^2H}{\tau_{\mathrm{diss}}} \lesssim 2\times10^{22}\,\mathrm{W} \approx 5\times10^{-5}L_\odot
\end{equation}
corresponding to a value of $\beta \sim 10^{-8}$\,m$^{-1}\ll R_{\rm s}^{-1}$.

We notice that the dissipation follows the latitudinal dependence of the $\alpha$ effect, which is concentrated toward the high latitudes, while it vanishes at the equator where $\alpha$ changes sign (cf. Eq.~\ref{beta_exp}). Therefore, in our simple model, the power dissipated per unit surface area varies with the colatitude proportionally to $\cos^{2} \theta$ and is maximum at the poles. An examination of available DAHe spectra demonstrates that the $\pm\sigma$ components of H$\alpha$ do not exhibit splitting variations consistent with those expected for a magnetic dipole.  Such changes would be on the order of 20\% because the observable field is integrated over the visible hemisphere, which averages the factor-of-two changes from pole to equator in a magnetic dipole.  However, only GD\,356 has data with full spin phase coverage, showing that its line splitting, coming from the integrated field of the visible hemisphere, hardly varies with rotation.

Three other DAHe stars show stronger variations, providing information about the distribution of field strength and emission regions over their surfaces. These are SDSS\,J1252$-$0234 \citep{Reding2020}, whose emission lines diminish into an absorption feature, then reappear as the star rotates; SDSS\,J1219+4715 \citep{Gaensicke2020}, whose emission strength varies, but not field strength; and WD\,J1616+5410 \citep{Manser2023}, for which emission strength and field strength may vary. These data indicate that the emission region does not fully cover the star and is likely concentrated in a patch \citep{Ferrario1997}. Also, the fields revealed in emission generally do not seem to show much broadening in their $\sigma$ components, which hints that, except for WD\,J1616+5410, the field strength in the emission regions may be uniform. Altogether, these observations may indicate that the assumption of a dipolar field could be a convenient simplification from a theoretical point of view, but not necessarily in agreement with observations.  Detailed models for these DAHe stars, from repeated spectropolarimetric measurements made through their rotation cycles, could provide insights into the validity of this model.

Our proposed mechanism is consistent with the lack of coronal and transition region emission in DAHe stars. To generate a corona, it is necessary to dissipate electric currents at high temperatures of the order of $10^{6}$\,K. However, an effective dissipation at temperatures of $\sim10^{6}$\,K is forbidden by the dramatic decrease in the magnetic diffusivity $\eta \propto T^{-3/2}$, which becomes $\sim 10^{-3}-10^{-4}$ times smaller than in the chromosphere. The rapid decrease of the electric resistivity with temperature serves as a ``thermostat'' which keeps most of the atmospheric plasma at $\sim10^{4}$\,K. We note that any plasma at $\sim 10^{5}$\,K (typical of the solar or stellar transition regions) is mainly heated by conduction from a hotter overlying corona, so we do not expect to observe extreme ultraviolet line emission \edit{if a corona is absent.  Furthermore, DAHe stars should have pure hydrogen atmospheres and thus lack any atomic transitions beyond the Lyman series.}

\section{Summary, implications, and future work}
\label{conclusions}

We have proposed that Ohmic heating from dynamo-driven currents is responsible for the DAHe phenomenon, ultimately caused by \edit{carbon--oxygen crystallization and distillation of $^{22}$Ne.  We argue that the high concentrations of $^{22}$Ne needed to drive the required dynamo can be naturally accounted for if DAHe white dwarfs are merger remnants.  In this picture, DAHe stars may be a lower-mass analogue of ultramassive Q branch white dwarfs whose observed cooling delays are likely also caused by $^{22}$Ne distillation.  However, we refrain from advocating a particular merger scenario for creating the required $^{22}\mathrm{Ne}$ abundances.} 

\edit{The main points of the DAHe white dwarf modeling can be summarized as follows:  
\begin{enumerate}[(i)]
\item{$^{22}$Ne distillation drives dynamos that result in strong magnetic fields.}
\item{Atmospheric heating is maximal at the magnetic poles, and zero at the equator.}
\item{Coronae are never produced owing to the temperature dependence of the resistivity.}
\item{Misaligned magnetic and spin axes will cause the chromospheric regions to rotate in and out of view.}
\item{A merger origin that enhances $^{22}$Ne is consistent with their relatively high masses and fast spins.}
\item{$^{22}$Ne distillation should facilitate a cooling delay, consistent with the observed clustering in the HR diagram.}
\end{enumerate}
}

\edit{The mechanism in this work requires a relatively high $^{22}\mathrm{Ne}$ abundance, but is agnostic as to its origin. A merger scenario has been invoked to address the overdensity of Q branch white dwarfs \citep[e.g.,][]{Shenetal23,Bedardetal24}, and merits further investigation in the context of the DAHe phenomenon.  Depending on the specific merger scenario, it might be expected that DAHe stars lack close, unevolved companions, similar to isolated magnetic white dwarfs suspected to have similar origins \citep{liebert2005,Tout2008,Garcia-Berro2012}.}

\edit{Furthermore, a key corollary of this model is that white dwarfs on the Q branch that distill $^{22}$Ne should also be magnetic and produce chromospheres.  Notably, there are two massive DQe white dwarfs known, both of which have emission lines detected only in the ultraviolet: both G227-5 and G35-26 exhibit emission lines of C\,{\sc ii} and O\,{\sc i}, where the latter star shows additional emission species \citep[e.g.\ N\,{\sc i}, Mg\,{\sc ii}, Si\,{\sc ii}][]{Provencaletal05}. {While neither star is currently known to be magnetic, 
hot DQ stars as a population are suspected stellar mergers} \citep{Dunlapetal15,Coutuetal19}, and possibly consistent with belonging to the delayed cooling population of the Q branch.  A modest number of hot DQ stars have been reported to be magnetic, fast rotators, but definitive studies and interpretations are lacking \citep{Dufouretal13,Lawrieetal13,Williamsetal16}. {As predicted, the two aforementioned DQe stars lie within the Q branch of the HR diagram. }}

\edit{One implication of our model is that DAHe white dwarfs should experience substantial cooling delays for many gigayears, depending on the abundance of $^{22}$Ne in their cores \citep{Camisassaetal21,Blouinetal21,Bedardetal24}.  While one might expect to find some ancient DAHe stars based on kinematics, the known members are a thin disk population with $\langle v_{\rm tan} \rangle=27\pm13$\,km\,s$^{-1}$.  From an observational perspective, this is a direct consequence of the {\em Gaia} magnitude limit, and the shared $M_G=13.2$\,mag of the DAHe population, restricting their detection to within roughly 150~pc, where thick disc and halo stars are rare \citep{Bensbyetal14,Zubiauretal24}.  From a theoretical viewpoint, any thick disc or halo mergers may have insufficient metallicities to produce the $^{22}$Ne necessary for distillation and stalled cooling \citep{Salarisetal24}.  It is therefore possible that few high-velocity members will be found among DAHe white dwarfs, at least until a sufficiently large sample is available.}

An interesting point to address in future investigations is the interaction between the atmospheric currents considered in the present model and hydromagnetic waves that can propagate along the same field lines \citep[e.g.][]{Musielak2005}. Such waves can be excited in the atmospheres of DAHe stars, because motions are still allowed along magnetic field lines \citep[cf.][]{Tremblayetal15}, and such motions will produce magneto-acoustic waves whose propagation will not be confined to the field lines. Such waves may produce shocks, heating the local atmosphere effectively.  Moreover, not all horizontal motions will be suppressed---some may survive and excite Alfv\'en waves which, although difficult to damp, can couple to magneto-acoustic waves and contribute to heating. Radiative damping effects, which are strong for pure acoustic waves in white dwarf atmospheres, would be significantly reduced for magneto-acoustic waves. On the other hand, they are absent for Alfv\'en waves because they are purely transverse magnetic modes that produce no heating of the plasma by compression. Therefore, the propagation of Alfv\'en waves cannot directly modify the plasma radiation.

Several aspects of the model we have introduced are addressed in a mainly qualitative way, although making use of the results of previous quantitative studies of stellar dynamos. A quantitative, mean field dynamo model based on numerical calculations can represent the next step of an investigation. Another topic to be addressed in future investigations is the inclusion of the proposed Ohmic heating in radiative transfer models of white dwarf atmospheres in order to provide quantitative predictions of the expected fluxes both in the continuum and in the spectral lines. Those predictions will hopefully allow a detailed test of the model proposed in the present study by means of a quantitative comparison with the observations.

\begin{acknowledgements}

The authors gratefully acknowledge an insightful and inspiring report by the referee, J.~Isern, who suggested to explore the $^{22}$Ne distillation as a potential source of energy to generate turbulent motions in white dwarfs. Moreover, the authors acknowledge highly valuable conversations and ideas shared by Z.~E.~Musielak and M.~R~Schreiber.  They are also grateful to J.~Fuller for useful comments, and to A.~B\'edard and M.~Hollands for clarifications on evolutionary model grids. This research was supported by the Munich Institute for Astro-, Particle and BioPhysics (MIAPbP) which is funded by the Deutsche Forschungsgemeinschaft under Germany's Excellence Strategy EXC~2094$-$390783311. AFL gratefully acknowledges support from the 2023 INAF Program for Fundamental Astrophysics through the project entitled "Unveiling the magnetic side of stars" (P.I. Dr.~A.~Bonanno) and from the European Union – NextGenerationEU RRF M4C2 1.1 n: 2022HY2NSX project "CHRONOS: adjusting the clock(s) to unveil the CHRONO-chemo-dynamical Structure of the Galaxy” (PI: Dr.~S.~Cassisi). NZR acknowledges support from the National Science Foundation Graduate Research Fellowship under Grant No.\ DGE$‐$1745301.

\end{acknowledgements}

\bibliographystyle{aa} 
\bibliography{Lanzaetal} 

\appendix
\section{Size and velocity of \edit{buoyant} flakes}
\label{Appendix_A}
\edit{When white dwarfs reach a central temperature equal to that of the crystallization phase transition, the formation of buoyant oxygen-rich crystals depleted of $^{22}$Ne occurs at their centres \citep[cf.][]{Isernetal91,Blouinetal21}.} The formation of solid flakes requires the diffusion of the crystallization latent heat to allow the fluid to become solid. If we indicate the size of a flake with $f$, the timescale for heat diffusion is $t_{\rm h} \sim f^{2}/\lambda$, where $\lambda$ is the thermal diffusivity. If a flake ascends with a terminal velocity $\varv_{\rm f}$, its characteristic dynamical timescale is of the order of $t_{\rm dyn} \sim 2f/\varv_{\rm f}$. We assume that the size of a flake and its terminal velocity are constrained by $t_{\rm h} \sim t_{\rm dyn}$ that gives the equation
\begin{equation}
    \frac{f^{2}}{\lambda} \sim \frac{2f}{\varv_{\rm f}}. 
    \label{eqf1}
\end{equation}
The terminal velocity $\varv_{\rm f}$ was estimated by \citet{Mochkovitch83} assuming the Stokes law, that is valid for a Reynolds number of order unity. As a matter of fact, it will be shown that the Reynolds number is remarkably larger than the unity, so we adopt the aerodynamic drag to describe the interaction between a rising flake and the surrounding fluid. By equating the drag to the buoyancy acting on a flake, we obtain its terminal velocity as 
\begin{equation}
\varv_{\rm f} \sim \left[ C_{\rm D}^{-1} f g \left( \frac{\Delta \rho}{\rho} \right) \right]^{1/2},
\label{eqf2}
\end{equation}
where $C_{\rm D}$ is the coefficient of aerodynamic drag that we assume to be unity for simplicity, $g$ is the acceleration of gravity, and $\Delta \rho /\rho$ is the relative density difference between a solid flake and the surrounding fluid (in absolute value), \edit{which we take to be in the range $0.01$--$0.1$ \citep[see][Fig.~2]{Blouinetal21}}. Making use of Eqs.~\eqref{eqf1} and~\eqref{eqf2}, we obtain
\begin{equation}
    f \sim (2\lambda)^{2/3} \left[ g \left(\frac{\Delta \rho}{\rho} \right) \right]^{-1/3} 
    \label{eqf3}
\end{equation}
and 
\begin{equation}
\varv_{\rm f} \sim \left[ f g \left( \frac{\Delta \rho}{\rho}\right)\right]^{1/2}. 
\label{eqf4}
\end{equation}
The thermal diffusivity $\lambda$ is a function of the electron thermal conductivity $\kappa_{\rm cond}$ \citep[][]{HubbardLampe69, Cassisietal07} according to 
\begin{equation}
\lambda = \frac{\kappa_{\rm cond}}{c_{\rm p} \, \rho},
\end{equation}
where $\rho$ is the plasma density and $c_{\rm p}$ is the specific heat at constant pressure that we assume to be approximately equal to $3k_{\rm b}$ per unit particle, where $k_{\rm b}$ is the Boltzmann constant, an expression appropriate at solidification neglecting the contribution of the electrons that is $\approx 25$\% of that of the ions \citep[cf.\ Sect.~37.3 of][]{Kippenhahnetal13}. Given that this treatment is only a rough approximation, we do not increase its uncertainty if we compute $\kappa_{\rm cond}$ from its ratio to the electric conductivity, that is a simple function of the temperature \citep[cf.][]{Itohetal08}, and adopt the electric conductivity for the interior of a white dwarf as estimated by \citet{Isern2017}. With this simplification,  we obtain $\lambda  \sim  10^{-3}$\,m$^{2}$\,s$^{-1}$ for a solidification temperature of $6\times 10^{6}$\,K \edit{as given by \citet{Castro-Tapiaetal24} for a white dwarf mass of 0.8\,M$_{\odot}$}. 
For an acceleration of gravity $g = 10^{7}$\,m\,s$^{-2}$, from Eqs.~\eqref{eqf3} and~\eqref{eqf4}, \edit{we obtain $f \sim (1.6-3.5) \times 10^{-4}$\,m and $\varv_{\rm f} \sim (6-12)$\,m\,s$^{-1}$.} The Reynolds number is $ f\varv_{\rm f}/ \nu \approx 670$ for a kinematic viscosity $\nu \sim 3 \times 10^{-6}$\,m$^{2}$\,s$^{-1}$ as estimated by \citet{Isern2017} \edit{and is independent of $\Delta \rho/\rho$}. A posteriori, such a value justifies the adoption of the aerodynamic drag formula instead of the Stokes formula. More importantly, it indicates that the wake past a rising solid flake is turbulent, thus providing a turbulent velocity field much faster than predicted by compositional convection according to \citet{Fuentesetal23}.  

\section{Diffusion of the magnetic field}
\label{app_b_diff}

Considering Eq.~\eqref{induction_eq}, and that the magnetic field is solenoidal, we obtain the equation ruling the diffusion of the field
\begin{equation}
\frac{\partial {\vec B}}{\partial t} = -\nabla \eta \times (\nabla \times {\vec B}) + \eta \nabla^{2} {\vec B}.
\label{eq_a1}
\end{equation}
Substituting the force-free relationship $\beta {\vec B} = \nabla \times {\vec B}$ into Eq.~\eqref{eq_a1}, we obtain
\begin{equation}
\frac{\partial {\vec B}}{\partial t} = -(\beta \nabla \eta) \times {\vec B} + \eta \nabla^{2} {\vec B}.
\label{fff_diff}
\end{equation}
The geometric interpretation of Eq.~\eqref{fff_diff} is that the time change of a force-free magnetic field has two components. The first corresponds to a slow rotation of the instantaneous field $\vec B$ around the local vector $-\beta \nabla \eta$, while the second represents the slow diffusion of the field itself. The fact that the first term produces a rotation of the vector $\vec B$, follows from the constancy of the modulus of $\vec B$ in the corresponding transformation. Specifically, the time variation of the square of the modulus of the magnetic field is
\begin{equation}
    \frac{\partial B^{2}}{\partial t} = 2{\vec B} \cdot \frac{\partial {\vec B}}{\partial t} = -2{\vec B}\cdot (\beta \nabla \eta) \times {\vec B}=0,
\end{equation} 
which vanishes because two vectors in the triple product are parallel.

We denote the length scale of the variation in $\eta$ as $L_{\eta}$, and that in the magnetic field as $L_{\rm B}$. Therefore, the order of magnitude ratio between the two components is, 
\begin{equation}
\frac{|(\beta \nabla \eta ) \times {\vec B}|}{|\eta \nabla^{2} {\vec B}|} \sim \beta L_{B}^{2} L_{\eta}^{-1}.
\label{oom}
\end{equation}
Assuming that $L_{\rm B} \sim L_{\eta}\ga 10^{5}$\,m, and $\beta \sim 3\Omega/\varv_{\rm t} \sim 3 \times 10^{-3}$\,m$^{-1}$ for the reference white dwarf, the numerator in Eq.~\eqref{oom} exceeds the denominator by a factor $\ga 300$. Since we are describing the process of field diffusion in the interior of the white dwarf, where electron degeneracy warrants a high conductivity, the value of $\beta$ is not limited by the dissipation of electric currents in the photosphere. \edit{In this regime, the magnetic field variation is dominated by the slow local rotation of the field, which does not change its topology, provided that the length scale over which the field varies is significantly smaller than the length scale over which the vector field $\beta \nabla \eta$ varies. We conclude that the field remains approximately force-free during its slow propagation towards the outer layers of the star.} We notice that the vector $\beta \nabla \eta$ has a latitudinal component as a consequence of the dependence of $\eta$ (and $\beta$) on the latitude in a rotating star. Therefore, the rotation around $\beta \nabla \eta$ produces a slow variation in the radial distribution of $\vec B$ that allows the field to move towards the surface of the star.

\section{Magnetic diffusivity in a white dwarf atmosphere}
\label{magn_diff}
{Magnetic diffusivity in a white dwarf atmosphere is given by $\eta = (\mu \sigma)^{-1}$, where $\mu$ is the magnetic permeability of the plasma, assumed equal to that in the vacuum, and $\sigma$ the electric conductivity. In a magnetized plasma, the magnetic field produces an anisotropic electric conductivity \citep[cf.\ Sect.~2.1.3 of][]{Priest1984} with a complex dependence on the densities of electrons, ions, and neutral particles as well as on the gradient of the electron pressure. However, if the current flows parallel to the magnetic field, as in our model, and the electron pressure gradient is negligible, the expression for the electric conductivity greatly simplifies as
\begin{equation}
    \sigma = \frac{n_{\rm e} e^{2}}{m_{\rm e} (\tau_{\rm ei}^{-1} + \tau_{\rm en}^{-1})},
    \label{electron_cond}
\end{equation}
where $n_{\rm e}$ is the electron density, $e$ the electron charge, $m_{\rm e}$ the electron mass, $\tau_{\rm ei}$ the typical electron-ion collision time interval, and $\tau_{\rm en}$ the typical electron-neutral collision time interval. For a fully ionized plasma, $\tau_{\rm en}^{-1}=0$, while according to Spitzer
\begin{equation}
\tau_{\rm ei} = 0.266 \times 10^{6} \frac{T^{3/2}}{n_{\rm e} \ln \Lambda} \mbox{ s},
\end{equation}
where $T$ is the temperature in K and $n_{\rm e}$ is measured in m$^{-3}$. The Coulomb logarithm is generally a slowly varying function of $T$ and $n_{\rm e}$ of the order of $5-20$ in the density and temperature regimes of white dwarf atmospheres, given the expression for the magnetic diffusivity in Eq.~\eqref{eta_value}. When the hydrogen plasma is partially ionized with a neutral density of $n_{\rm n}$, the magnetic diffusivity is increased by the factor $(1+ \tau_{\rm ei}/\tau_{\rm en})$. In the photosphere of a white dwarf at temperatures of a few thousand K, $n_{\rm n} > n_{\rm e}$, $\tau_{\rm en} < \tau_{\rm ei}$, thus increasing the Ohmic heating with respect to the value computed with the Spitzer formula \eqref{eta_value}. Therefore, the estimate of the Ohmic heating is actually a lower value. However, as soon as the temperature is increased by the Ohmic dissipation, a larger fraction of hydrogen (and helium) ionizes making the actual value of $\eta$ closer to the approximation. 

Electron degeneracy is usually negligible in the photosphere of white dwarfs \citep[cf.\ Sect.~37 of ][]{Kippenhahnetal13}. The effect of electron degeneracy in a fully ionized plasma is to modify $\tau_{\rm ei}$ and introduce the free electron effective mass $m_{\rm R}$ in place of $m_{\rm e}$ in Eq.~\eqref{electron_cond} \citep[cf.\ Eq.~(1) of][]{NandkumarPethick84}. In the density and temperature regimes characteristic of white dwarf photospheres, we do not expect any relevant modification of the Spitzer formula because the small electron degeneracy does not significantly modify the electron-ion collision time, while $m_{\rm R} \simeq m_{\rm e}$ because relativistic effects are negligible. }

\end{document}